\documentstyle[12pt]{article}

\def\a{\alpha}

\def\s{\sigma}
\def\sb{\bar{\sigma}}
\def\th{\theta}
\def\thb{\bar{\theta}}

\def\ud{\underline}

\def\G{\Gamma}

\newskip\humongous \humongous=0pt plus 1000pt minus 1000pt

\newif\ifdtup

\def\ket#1{\left| #1\right\rangle}

\def\beq{\begin{equation}}
\def\eeq{\end{equation}}

\def\beqn{\begin{eqnarray}}
\def\eeqn{\end{eqnarray}}
\relax

\def\G2{{\; \rm GeV/}c^2}
\def\G{\; \rm GeV}

\def\dotx{\dotx{\dot\overline{x}}}

\relax

\newcommand{\trans}[1]{{t\  \atop \displaystyle \:}\!\!\!\! {#1}}

\begin{document}
\begin{titlepage}
\begin{flushright}
       {\normalsize  OU-HET 305 \\  hep-th/9810237 \\
           version  to appear in Modern Physics Letters A ,  October, 1998}
\end{flushright}
%
\begin{center}
  {\large \bf  Nonabelian Berry Phase,  Yang-Mills Instanton  \\
 and $USp(2k)$ Matrix Model }\footnote{This work is supported in part
 by the Grant-in-Aid  for Scientific Research (10640268) and
 Scientific Research Fund (97319)
from the Ministry of Education, Science and Culture, Japan.}

\vfill
          {\bf B.~Chen}  \\
         {\bf H.~Itoyama}  \\
            and \\
         {\bf H.~ Kihara}\\
        Department of Physics,\\
        Graduate School of Science, Osaka University,\\
        Toyonaka, Osaka, 560 Japan\\
\end{center}
\vfill
\begin{abstract}
   The nonabelian Berry phase is computed in the $T$ dualized quantum
 mechanics obtained from the $USp(2k)$ matrix model. Integrating the fermions,
 we find  that each of the spacetime points $X_{\nu}^{(i)}$ is equipped with
  a pair of $su(2)$ Lie algebra valued pointlike singularities located
 at a distance $m_{(f)}$ from the orientifold surface.
 On a four dimensional paraboloid embedded in the five dimensional
 Euclidean  space, these singularities are recognized as
  the BPST instantons.

\end{abstract}
\vfill
\end{titlepage}

  Roles of instanton configurations in nonabelian gauge theory 
 have been thoroughly investigated in the literature in various circumstances.
 More recently, the Yang-Mills instanton has turned out to play an
 interesting role
 in (super)string theory as well  with the advent of $D$ branes \cite{Pol}.
  In ref. \cite{Witten},  Witten has argued that
 the zero size limit of the  Yang-Mills instanton is identified with
 the five brane of type $I$ superstring theory. Equivalence of
  branes and their bound states with
 gauge theory instantons appears to be a general feature of superstrings
  \cite{Douglas} and should be demonstrated fully with ample examples. 
 In this letter, we will find that a particular formulation of  unified  
 superstring theory via the matrix model also exhibits  the BPST
 instanton \cite{BPST}.  A notable feature  is that this configuration
 emerges naturally after the integrations of fermions.

  There are several matrix models \cite{BFSS,Mhetero,IKKT,IT1,IT2} proposed
 for unified superstring theory up to now, which embody the notion of
 noncommuting coordinates \cite{NCC}.
  The model we will adopt below  is the reduced  ({\it i.e.}
   zero-dimensional ) model based on the $USp$ Lie
 algebra descending from Type $I$ superstrings \cite{IT1,IT2}.
  We find that the effective action which depends on the path in the space
 of spacetime points $X_{M}^{(i)}$ (which are dynamical variables)
 sitting at the diagonal entries of the bosonic matrices contains
  a coupling to the instanton background. This background
 emerges from integrations of the fermionic degrees of freedom
 \footnote{ See \cite{fermi,IM} for similar computation.}.      
 The nonabelian Berry phase \cite{NAB}
  ( see also \cite{Berry,Stone}) plays a decisive role in this phenomenon.
  The degrees of freedom belonging to the fundamental representation
  will turn out to give surviving contributions after the cancellations due to
 the symmetry of  the roots and that of the weights are taken into account.

  Let us briefly review  the construction of the reduced $USp(2k)$ matrix
 model.  The action consists of the three parts  and can be written 
 by borrowing
  the $d=4$ superfield notation and dropping the spacetime dependence.
  We denote by $S_{{\rm vec}}$  the action  for the vector multiplet of
  $d=4, {\cal N}=2$  $USp(2k)$ supersymmetric  gauge theory
 and by $S_{{\rm asym},({\rm fund})}$  that of the hypermultiplet
  belonging to the antisymmetric ( fundamental ) representation.  
  ( See \cite{IT2} for detail.) The action, which is denoted by $S_{0}$ in the
 absence of $n_{f}$ of the fundamental hypermultiplets, is expressible as
\beqn
   S_{0} &=&  S_{{\cal N}=1}^{d=10}(
 \hat{\rho}_{b\mp}\ud{v}_{M}, \hat{\rho}_{f\mp}\ud{\Psi}  ) \;\;,
\label{eq:ours=10dSYM} \\
   S^{d=10}_{{\cal N}=1}(\ud{v}_{M},  \ud{\Psi} )  &=&
  \frac{1}{g^{2}} Tr \left( \frac{1}{4} \left[\ud{v}_{M},
 \ud{v}_{N} \right]
\left[ \ud{v}^{M}, \ud{v}^{N} \right] - \frac{1}{2}
\bar{\ud{\Psi}} \Gamma^{M} \left[ \ud{v}_{M},
\ud{\Psi} \right] \right) \;\;.
\eeqn
  This action can be understood as the projection from
 the type $IIB$ matrix model. The attendant projector
\beqn
 \hat{\rho}_{\mp} \bullet  \equiv \frac{1}{2} \left( \bullet \mp F^{-1}
 \bullet^{t}  F \right) \;\;
\eeqn
  takes  any $U(2k)$ matrix (denoted by a symbol with an underline)
  into the matrix lying in the adjoint
representation  of $USp(2k)$  and that in the antisymmetric
representation respectively.  The symbol $\hat{\rho}_{b\mp}$ is a matrix
 with Lorentz indices
and  $\hat{\rho}_{f\mp}$ is a matrix with spinor indices:
\beqn
 \hat{\rho}_{b\mp}
 &=& diag
(\hat{\rho}_{-},\hat{\rho}_{-},\hat{\rho}_{-},
 \hat{\rho}_{-},\hat{\rho}_{-},\hat{\rho}_{+},
 \hat{\rho}_{+},\hat{\rho}_{-}, \hat{\rho}_{+},\hat{\rho}_{+} )
\nonumber \\
 \hat{\rho}_{f\mp}   &=&
\hat{\rho}_{-} 1_{(4)} \otimes
\left( \begin{array}{cccc}
        1_{(2)}& & & \\
               & 0 & & \\
               &   & 1_{(2)} & \\
               &   &         &0
        \end{array}
\right)
+
\hat{\rho}_{+} 1_{(4)} \otimes
\left( \begin{array}{cccc}
        0& & & \\
               & 1_{(2)} & & \\
               &   & 0 & \\
               &   &         & 1_{(2)}
        \end{array}
\right)
 \;\;.
\label{eq:projectors of b f}
\eeqn
 The third part  $S_{{\rm fund}}$ reads
\beqn
 S_{{\rm fund}} &=&  \frac{1}{g^{2}} \sum_{f=1}^{n_f}
\left[
  \int d^2 \th d^2 \thb
\left( Q_{(f)}^{* \;i} \left( e^{2V} \right)_i^{\;\;j} Q_{(f) \; j}
+  \tilde{Q}_{(f)}^{ i} \left( e^{-2V} \right)_i^{\;\;j}
        \tilde{Q}_{(f) \; j}^{*} \right)
\right.   \nonumber  \\
& &
\qquad
 +
\left.
\left\{
\int d^2 \th
\left(
        m_{(f)} \tilde{Q}_{(f)}^{\;\;\;\; i}  Q_{(f) \;i}
        + \sqrt{2} \tilde{Q}_{(f)}^{\;\;\;\; i}
                 \left( \Phi \right)_i^{\;\;j} Q_{(f) \;j}
\right)
+ h.c.
\right\}
\right]  \;\;, \\
Q_{i} &=&  Q_{i} + \sqrt{2} \th \psi_{Q \; i} + \th \th
 F_{Q \; i} \;\;.
\eeqn
 This part is designed to create an open string sector.

  It has been demonstrated in \cite{IT1,IT2} that the model is uniquely
 selected by the three requirements: 1)having eight dynamical and eight
 kinematical supercharges, 2)obtained by an appropriate projection
 from the $IIB$ matrix model and an addition of the degrees of freedom
 corresponding
 to open strings, 3)nonorientable. The last requirement selects the $usp$ Lie
 algebra  both  from the planar diagram analysis \cite{IT2}
  and more clearly from the structure of the large $k$ Schwinger-Dyson
 equations
 or equivalently that of the closed and open loop equations \cite{ITsu}.
   This latter analysis brings at the same time
 the $SO(2n_{f})$ Chan-Paton factor coming from the flavor symmetry.
  Reflecting  eq.~(\ref{eq:projectors of b f}), the classical vacuum is
 broken by $Z_{2}$ in the six adjoint directions through
 orientifold projections. The mass $m_{(f)}$ denotes the distance of
 $D$-objects from orientifold surfaces, which
 is the case in the representation of strings as effective worldvolume
 gauge theory.  This will be seen to hold in the present discussion
  based on matrices as well.
  Letting $k$ to infinity enables us to construct matrix $T$ duality
 transformation via the recipe of \cite{WT}. 
  Via this transformation, the perturbative properties of strings
 built from matrices can be represented as the worldvolume
  $USp$ gauge theories in various dimensions.   The consistency
  with the literature has been checked  in  some cases \cite{IT2}.

  Let us construct the above mentioned  effective action which depends on
 the paths $\{ \Gamma_{A}^{(R)} \}$ in the parameter space
  labelled by the five sets of the adjoint spacetime points
$X_{\nu} = diag(X_{\nu}^{(1)}, \cdots X_{\nu}^{(k)},$$ -X_{\nu}^{(1)},
 \cdots -X_{\nu}^{(k)})$ $\nu =1,2,3,4,7$.
  This can be accomplished by utilizing the representation  of the model
 as $T$ dualized quantum mechanics \cite{0T}. ( We choose $v_{0}=0$ gauge):
\beqn
\label{eq:adprocess}
Z \left[ X_{\nu}; m_{(f)}, \{\Gamma_{A}^{(R)} \},
  \{ \sigma_{i A}^{(R)} \},  \{ \sigma_{f A}^{(R)} \}
 \right] =  \int \left[ D \tilde{v}_{M} \right]    \prod_{f=1}^{n_{f}}
\left[ D Q_{(f)}\right] \left[ D Q_{(f)}^{*}\right]
  \left[D {\tilde{Q}}_{(f)} \right]
\left[ D \tilde{Q}_{(f)}^{*} \right]  e^{iS_{B}}  \nonumber \\
 \lim_{T \rightarrow \infty} \langle  \Psi ; \{ \sigma_{i A}^{(R)} \}  
  \mid P \exp
 \left[ -i  \int_{0}^{T} dt H_{fermion}(t) \right]
 \mid \Psi ; \{ \sigma_{f A}^{(R)} \}  \rangle \;. 
\eeqn
  Here $v_{M} = X_{M}+ \tilde{v}_{M}$ and $S^{B}$ is the pat of $S$
which does not contain any fermion. For simplicity, we have set the dependence
of the remaining antisymmetric spacetime points $X_{I}=  diag(X_{I}^{(1)},
 \cdots X_{I}^{(k)},$$ X_{I}^{(1)}, \cdots X_{I}^{(k)})$ $\;I= 5,6,8,9$ 
  to zero \footnote{ This appears to be a valid approximation as worldvolume
 gauge theories in various dimensions obtained as matrix $T$ duals from
  this model  have string interpretation on their Coulomb phase.
   See, for example, \cite{wveg}.}.
Due to this choice,
 the operator $H_{fermion}$ becomes a direct sum of the three
 Hamiltonians $H_{{\rm fund}},H_{{\rm adj}}$ and $H_{{\rm asym}}$
  respectively obtained from the fermionic part of
$ S_{{\rm fund}}, S_{{\rm adj}}$ and $S_{{\rm asym}}$ after
 $T$ duality \footnote{ Complete analysis which includes the antisymmetic
 spacetime points is now in progress by us \cite{CIK2}. This will further
 clarify the physical picture of the nonabelian Berry phase in the
  $USp(2k)$ matrix model.}.
Their $t$ dependence comes from that of $X_{\nu}$  which acts as external
parameters on the Hilbert space of fermions. 
The quantity (\ref{eq:adprocess}) was studied in \cite{IM}, ignoring the
degeneracy of the ket vector $\mid \Psi \rangle$ which is an adiabatic
eigenstate of the system. Here we consider a set of degenerate adiabatic
eigenstates. The degeneracy of the initial state and that of the final one
are respectively specified by  a set of labels  $\{ \sigma_{i A}^{(R)} \}$
and $\{ \sigma_{f A}^{(R)} \}$, where the indices $A$ and $(R)$  specify
 the species of fermions.

  Let $R = {\rm fund, adj, antisym}$. We denote by $e_{(R)}^{(A)}$
the standard eigenbases belonging to the roots of $sp(2k)$ and the weights
 of the fundamental representation and those of the antisymmetric
representation respectively.
Let us expand the two component fermions as
\beqn
\label{eq:expand}
 \psi^{(R)} = \sum_{A}^{N_{(R)}} b_{A}^{(R)}
 e_{(R)}^{(A)} /\sqrt{2} \;,\;\;
\bar{\psi}^{(R)} = \sum_{A}^{N_{(R)}} \bar{b}_{A}^{(R)}
 e_{(R)}^{(A) \dagger}/ \sqrt{2} \;\;\;,
\eeqn
  where  $N_{({\rm adj})}= 2k^{2} +k$,
  $N_{({\rm antisym})}= 2k^{2} -k$ and  $N_{({\rm fund})}= 2k$.
  We find that all of the three Hamiltonians
 $H_{{\rm fund}},H_{{\rm adj}}$ and $H_{{\rm asym}}$
 are expressible in terms of  a generic one
\beqn
\label{eq:abelian}
  g^{2} H_{0} \left(X_{\ell}, \Phi, \Phi^{*}; (R), A \right) =
-\bar{b}_{A\dot{\a}}^{(R)} {\sb}^{m\dot{\a}\a}X_{m}b_{A\a}^{(R)}
-d^{(R)\a}_{A} \s_{\a\dot{\a}}^{m}X_{m} \bar{d}_{A}^{(R) \dot{\a}}
+ \sqrt{2}  \Phi b^{(R) \a}_{A}  d_{A\a}^{(R)}  \nonumber \\
+\sqrt{2}\Phi^{*}\bar{b}_{A\dot{\a}}^{(R)}
  \bar{d}^{\dot{\a} (R) }_{A} \;\;
\eeqn
provided we replace the five parameters
\beq
X_{\ell},\;\;\; \Phi = \frac{X_4 + iX_7}{\sqrt{2}}, \;\;\;
\Phi^{*} = \frac{X_4 - iX_7}{\sqrt{2}} \;\;\; \ell =1,2,3
\eeq
by the appropriate ones. (See argument of ${\cal A}$ in
eq.~(\ref{eq:answer1}) below.)

The Berry connection appears in one or three particle
 state of $H_{0}$ \cite{IM} with respect to
 the Clifford vacuum $\ket{\Omega}$;
 $b^{\a}\ket{\Omega} = \bar{d}_{\dot{\a}}\ket{\Omega} = 0$.
  We suppress the labels $A$ and $(R)$ seen in
 eqs. (\ref{eq:expand}),(\ref{eq:abelian}) for a while.
  Let us write   $\mid \Psi  \rangle =  \left( h_{\a} d^{\a}
+ \bar{h}^{\dot{\a}} \bar{b}_{\dot{\a}}  \right) \mid \Omega \rangle $ and
 $\psi \equiv \left( h_{\a} , \bar{h}^{\dot{\a}} \right)^{t}$.
 The transition amplitude of an adiabatic process  reads
\beqn
\label{eq:formula}
 \lim_{T \rightarrow \infty} \langle \Psi \mid P \exp \left[ -i  
\int_{0}^{T}
 dt H_{0}(t) \right] \mid  \Psi \rangle
  =  \psi^{\dagger}
 P \exp \left[ -i \int_{0}^{\infty} E(t) dt
 + i \int_{\Gamma} d\gamma (X_{m}, \Phi, \Phi^{*})  \right]
 \psi\;.
\eeqn
Here $\Gamma$ is a closed path in the parameter space. The  
connection one-form  is
\beqn
  id\gamma (t)= -  \psi^{\dagger} (t) d \psi (t)
  \equiv   -i {\cal A} \;\;\;,
\eeqn
  which is in general matrix-valued.
 Let us consider for definiteness a set of two degenerate adiabatic
 eigenstates with positive energy, which is specified by an index
 $\sigma =1,4$.  Using the completeness
  $ {\displaystyle \sum_{\sigma =1,4} } \psi_{\sigma}
 \psi_{\sigma}^{\dagger} = {\bf 1}_{(2)}$,
   we find
\beq
\label{eq:factor}
\sum_{\sigma = 1,4}  \lim_{T \rightarrow \infty} \langle \Psi_{\sigma}
 \mid P \exp \left[ -i \int_{0}^{T}
 dt H_{0}(t) \right] \mid  \Psi_{\sigma} \rangle =
  tr  P \exp \left[ -i \int_{0}^{\infty} E(t) dt
 - i \int_{\Gamma} {\cal A}(X_{\nu}) \right] 
\eeq
 from eq. (\ref{eq:formula}).
  Here the trace is taken with respect to the two-dimensional subspace.

  Now the problem is to obtain the nonabelian $(su(2)$ Lie algebra valued)
  Berry connection associated with the first quantized hamiltonian
\begin{equation}
{\cal H} = \frac{R}{g^2} \sum_{\nu=1,2,3,4,7}  N^{\nu}  \Gamma_{\nu}\;, \;
   R \equiv \sqrt{(X^1)^2 +(X^2)^2 +(X^3)^2 +(X^4)^2 +(X^7)^2}\;,
 \; N^{\nu} \equiv \frac{X^{\nu}}{R} \;,
\end{equation}
  where $\Gamma_{\nu}$ are the five dimensional gamma matrices obeying
 the Clifford algebra and the explicit representation can be read off
  from eq.~(\ref{eq:abelian}).
  The projection operators are
\begin{equation}
P_{\pm} = \frac{1}{2} ({\bf 1}_4 \pm  N^{\nu} \Gamma_{\nu}) \;\;, \;\;
{P_{\pm}}^2 = P_{\pm} \;\;,\;\; 
 P_+^{\dag} = P_+\;\;, 
\end{equation}
   which satisfy
\begin{equation}
{\cal H} P_{\pm} = \pm \frac{R}{g^2} P_{\pm}\;\;.
\end{equation}
  Denoting by ${\bf e}_i$ $(i=1,2,3,4)$ the unit vector in the $i$-th
 direction,  we write a set of normalized eigenvectors belonging to
 the plus eigenvalue as
\begin{equation}
 \psi_i  = \frac{1}{{\cal N}_i} P_{+}{\bf e}_i \;\;.
\end{equation}
 Here, $i=1,4$ refer to the sections around the north pole  $X^{3}=R$ 
  while  $i=2,3$  to the ones around the south pole $X^{3}=-R$.
  The ${\cal N}_i$ are the normalization factors:
\begin{equation}
{\cal N} \equiv {\cal N}_1 = {\cal N}_4 = \sqrt{\frac{1+N^3}{2}}\;\;,\;\;
 {\cal N}^{\prime} \equiv{\cal N}_2 = {\cal N}_3 = \sqrt{\frac{1-N^3}{2}}\;\;.
\end{equation}
  We focus our attention on the sections near the north pole. The Berry
 connection  is
\begin{equation}
i{\cal A}  = \left( \begin{array}{l}
			  \psi_1 \\
			 \psi_4 
		\end{array} \right)
	 d (  \psi_1,  \:  \psi_4  )
 =  \left( \begin{array}{l}
			\trans{{\bf e}_1} \\
			\trans{{\bf e}_4}
		\end{array} \right)
    {\cal M} ( {\bf e}_1 \: {\bf e}_4 )\;\;,
\end{equation}
 where ${\cal M} \equiv \frac{1}{{\cal N} }P_+^{\dag} d 
 \frac{1}{{\cal N}}P_+ $.
  Introducing $C^{\mu \nu} \equiv (N^{\mu} dN^{\nu} - N^{\nu} dN^{\mu})$,
 we obtain
\begin{equation}
{\cal M}= \frac{1}{1 + N^3} \left( \frac{1}{2}dN^{\mu} \Gamma_{\mu} +
\frac{1}{4}C^{\mu \nu} \Gamma_{\mu} \Gamma_{\nu} -
 \frac{1}{1 + N^3} dN^3 P_+ \right) \;\;.
\end{equation}
   Working out  each entry of the $2\times2$ matrix ${\cal A}$,  we find,
  after some calculation,
\begin{equation}
\label{NABCAL} 
{\cal A}  \left( X^{i} \right) = \frac{R}{2(R + X^3)}
 \frac{1}{R^2} {\bf B} \cdot
 \mbox{\boldmath $\sigma$},
\end{equation}
where
\begin{equation}
{\bf B}   \equiv
 \left[
\begin{array}{l}
     B^1 \\
     B^2 \\
     B^3
\end{array}\right]   =
 \left[
\begin{array}{l}
X^7 dX^1 - X^1 dX^7 - X^2 dX^4 + X^4 dX^2 \\
X^1 dX^4 - X^4 dX^1 - X^2 dX^7 + X^7 dX^2 \\
X^4 dX^7 - X^7 dX^4 - X^2 dX^1 + X^1 dX^2
\end{array}\right]\;\;.
\end{equation}
  Observe that $R$ and $X^3$ appear  only in the overall scale factor.

To proceed further, we parametrize  $S^{3}$ of unit radius by the coordinates
\begin{equation}
Y^{\nu} \equiv \frac{1}{\sqrt{R^2 - (X^3)^2}} X^{\nu}\:\:,\:\:\:
 (\nu = 1,2,4,7)
\end{equation}
\begin{equation}
(Y^1)^2 + (Y^2)^2 + (Y^4)^2 + (Y^7)^2 = 1\;\;.
\end{equation}
  We find
\begin{equation}
\label{eq:B}
 \frac{1}{R^2 - (X^3)^2} {\bf B} = 
 {\bf Y} \times d{\bf Y} + {\bf Y} dY^2 - Y^2 d{\bf Y}\;\;,
\end{equation}
  where ${\bf Y} \equiv ( Y^4, Y^7, Y^1 )^{t}$.
  The $Y$ coordinates parametrize the $SU(2)$ group element as well:
\begin{equation}
T \equiv Y^2 {\bf 1}_2 + i {\bf Y} \cdot \mbox{\boldmath$\sigma$}\;\;,
\end{equation}
  from which  we can make the pure gauge configuration
\begin{equation}
\begin{array}{rcl}
dT T^{-1} 
 &=& (dY^2 {\bf 1}_2 + i d{\bf Y} \cdot \mbox{\boldmath$\sigma$})
(Y^2 {\bf 1}_2 - i {\bf Y} \cdot \mbox{\boldmath$\sigma$})\\
 &=& i (d{\bf Y} \times {\bf Y} + Y^2 d{\bf Y} -{\bf Y} dY^2) \cdot 
\mbox{\boldmath$\sigma$}.
\end{array}
\end{equation}
  Eq.(\ref{NABCAL}) and eq.(\ref{eq:B}) tell us
\begin{equation}
\label{eq:BPST}
{\cal A} \left( X^{\nu} \right) = p(R, X^{3}) dT T^{-1}\;\;.
\end{equation}
 The prefactor $p(R, X^{3})$ is of interest  and  can  be written as
\beq
 p(R, X^{3}) =  \frac{\tau^2}{\tau^2 + \lambda^2}  \;\; , \;\;\;
\tau = \sqrt{R^2 - (X^3)^2}\;\;,\;\;\;
\lambda = R + X^3.
\eeq
 The nonabelian connection ${\cal A}$ is in fact the instanton
 configuration of Belavin et. al \cite{BPST}.  The size of the instanton
 $\lambda$
 is not a ${\it bonafide}$ parameter in the model but is chosen to be
 the fifth
 coordinate in  the  five dimensional Euclidean space.  For fixed $\lambda$, 
 the four dimensional subspace embedded into  the ${\bf R}^{5}$  is a 
 paraboloid  wrapping the singularity. An observer on this recognizes
 the pointlike singularity  as  BPST instanton.  As $\lambda$ goes to zero,
 this paraboloid gets degenerated into an $SU(2)$ counterpart of the Dirac
 string connecting  the origin and the infinity.
     Note also that the prefactor  is written in terms of  the angle measured
 from the north pole as
\begin{equation}
p(\theta) = \frac{1}{2}(1- \cos \theta )\;\;,\;\;
N^3  \equiv R \cos \theta \;\;.
\end{equation}

 Returning to the expression (\ref{eq:adprocess}) and
 taking a sum over the labels $\sigma_{f A}^{(R)} = \sigma_{i A}^{(R)}$,
  we find that the second line is expressible as the product of the factors
\beqn
\label{eq:answer1}
 &&  Tr P \exp \left( -i \sum_{f=1}^{n_{f}} \sum_{A=1}^{2k}
 {\bf 1} \otimes \cdots \int_{\Gamma_{A,f}^{(fund)} } {\cal A}
\left[ {\bf w}^{A}\cdot {\bf X} _{\ell},\;
 \frac{ m_{(f)}}{\sqrt{2}}    +{\bf w}^{A} \cdot {\bf\Phi},\;
 \frac{ m_{(f)}}{\sqrt{2}} + {\bf w}^{A} \cdot {\bf \Phi }^{\dagger}
 \right] \cdots \otimes {\bf 1} \right) 
   \nonumber \\
  &&  Tr  P exp \left( -i \sum_{A=1}^{2k^{2}}
   {\bf 1} \otimes \cdots \int_{\Gamma_{A}^{(adj)} } 
  {\cal A} \left[ {\bf R}^{A} \cdot {\bf X}_{\ell},\;
 i {\bf R}^{A} \cdot {\bf \Phi},\;
  i {\bf R}^{A} \cdot {\bf \Phi}^{\dagger} \right]  \cdots
 \otimes {\bf 1} \right)  \nonumber \\  
 &&  Tr  P exp \left( 
 -i  \sum_{A=1}^{2k^2 - 2k} {\bf 1} \otimes \cdots 
 \int_{\Gamma_{A}^{(asym)} }
 {\cal A} \left[ {\bf w}_{{\rm asym}}^{A} \cdot {\bf X}_{\ell},\;
 {\bf w}_{{\rm asym}}^{A} \cdot {\bf \Phi},\;
  {\bf w}_{{\rm asym}}^{A} \cdot {\bf \Phi}^{\dagger} \right] \cdots
 \otimes {\bf 1} \right) \;\;. 
\eeqn
   We have included  the energy dependence seen in eq. (\ref{eq:factor})
 in $S_{B}$ as  this is perturbatively cancelled by the contribution from
 bosonic integration.  The symbols seen in the arguments are
\beqn
\{ \{  {\bf w}^{A} \mid  1 \leq A \leq 2k  \}\}
&=&
\{ \{   \pm {\bf e}^{(i)}
\; , 1 \leq i \leq k   \}\}   \;\; \nonumber \\
\{ \{ {\bf R}^{A} \mid 1 \leq A \leq 2k^{2} \}\}
&=&
\{ \{   \pm 2 {\bf e}^{(i)}, {\bf e}^{(i)}-{\bf e}^{(j)} ,
\pm \left({\bf e}^{(i)} +{\bf  e}^{(j)} \right)
\; 1 \leq i,j, \leq k   \}\}  \;\; \nonumber \\
\{ \{  {\bf w}_{ {\rm asym} }^{A} \mid  1 \leq A \leq 2k^{2} -2k   
\}\}
&=&
\{ \{    \pm \left( {\bf e}^{(i)} +{\bf  e}^{(j)}  \right) ,
{\bf e}^{(i)} - {\bf e}^{(j)},  \; 1 \leq i,j, \leq k   \}\} \;\;.
\eeqn	
  The second and the third lines
 are respectively the nonzero roots and the weights in the  
antisymmetric representation
 of $usp(2k)$. We have denoted by ${\bf e}^{(i)} \;(1 \leq i \leq  k)$
the orthonormal basis vectors of $k$-dimensional Euclidean space and
\beqn
{\bf X}_{\ell} = \sum_{i=1}^{k}  {\bf e}^{(i)} X_{\ell}^{(i)},\;
{\bf \Phi} =  \sum_{i=1}^{k}{\bf e}^{(i)}
\frac{X_{4}^{(i)} + iX_{7}^{(i)}}{\sqrt{2}}, \;
{\bf \Phi}^{\dagger}= \sum_{i=1}^{k}{\bf e}^{(i)}
\frac{X_{4}^{(i)} - iX_{7}^{(i)}}{\sqrt{2}}
\;\;\;.
\eeqn

  Unlike the case of the abelian Berry phase examined in \cite{IM}, the
second or the third line of eq. (\ref{eq:answer1}) do not quite collapse to
 unity.   Let us, however, exploit the symmetry of the roots and the weights
under ${\bf e}^{(i)} \leftrightarrow -{\bf e}^{(i)}$.
  Observe that, due to this symmetry,   we can pair
 the two-dimensional vector space associated with $ {\bf R}^{A}$
  (or ${\bf w}_{ {\rm asym} }^{A}$) and that with   $-{\bf R}^{A}$
 (or $-{\bf w}_{ {\rm asym} }^{A}$).  Let us symmetrize  the  tensor product
 of these two two-dimensional vector spaces. On this, the nonabelian
 Berry phase is reduced to the pure gauge configuration
\beq
 {\cal A}(X^{\nu})_{ \{i }^{ \;\;\{j } \delta_{k \} }^{\;\; \ell \} }
 + {\cal A}(-X^{\nu})_{ \{i }^{ \;\;\{j } \delta_{k \} }^{\;\; \ell \} }    
 = \left( dT T^{-1} \right)_{ \{i }^{ \;\;\{j }
 \delta_{k \} }^{\;\; \ell \} }   \;\;\;,
\eeq   
 and this can be gauged away.
  As for the first line of eq. (\ref{eq:answer1}), the mass terms prevent
 this from happening.

  After all these operations, eq. (\ref{eq:answer1}) becomes
\beqn
\label{eq:answer2}
   Tr P \exp \left( -i \sum_{f=1}^{n_{f}} \sum_{A=1}^{2k}
 {\bf 1} \otimes \cdots \int_{\Gamma_{A,f}^{(fund)} } {\cal A}
 \left[ {\bf w}^{A}\cdot {\bf X} _{\ell},\;
 \frac{m_{(f)}}{\sqrt{2}}    +{\bf w}^{A} \cdot {\bf\Phi},\;
 \frac{m_{(f)}}{\sqrt{2}} + {\bf w}^{A} \cdot {\bf \Phi }^{\dagger}
 \right]  \cdots \otimes {\bf 1} \right) \;.
\eeqn
  Note that in the antisymmetrized part of eq. (\ref{eq:answer1})
  the nonabelian Berry phase is  present generically in all three lines.
 It is present in the $IIB$ matrix case as well.

 We have a collection of pairs of Yang-Mills instanton configurations
which preserve a fraction of supersymmetries and
 which will presumably take their worldvolumes in four of the
antisymmetric directions representing ${\bf R}^{4}$ and time.
 The natural interpretation along the line of \cite{Witten} and
 \cite{Douglas} is that we have a collection of pairs
of $D0-D4$ bound state system.  This point will become clear
  in the full-fledged calculation which includes the
 antisymmetric spacetime points \cite{CIK2}.
 Each pair belongs to a different
 spacetime  point $X_{\nu}^{(i)}$ whose quantum mechanical average
 over all $i$ predict spacetime properties such as the size of the universe.
  In the classical consideration, they are located at a distance
 $m_{(f)}$ away from the orientifold plane.  This of course conforms to
 the picture already developed in \cite{IT2, ITsu}.
 A difference from the case of \cite{Witten} is that we have an observer
 dependent size of the instanton.
  It is more appropriate, however, to view this configuration as
  a pointlike nonabelian Yang monopole \cite{Yang} in
the  five dimensional space. This will be an eventual picture
 identified with  D-branes and their bound states.
 Another notable feature of our instanton/Yang monopole is that
  the $su(2)$ index originates from the spinor index of the
 fermions.  These points certainly deserve further studies.

\bigskip

 The authors thank Toshihiro Matsuo, Toshio Nakatsu and
 Asato Tsuchiya for helpful discussion on this subject.

\end{document}